\documentclass[aps,prb,twocolumn,groupedaddress,showpacs]{revtex4}

\usepackage{graphicx}
\begin{document}

\title{Field sweep rate dependence of the coercive field of single-molecule 
magnets: a classical approach with applications to the quantum regime}

\author{W. Wernsdorfer$^1$,  M. Murugesu$^2$, A.J. Tasiopoulos$^2$, and G. Christou$^2$}

%\email[]{wernsdor@grenoble.cnrs.fr}
%\thanks{}

\affiliation{
$^1$Lab. L. N\'eel, associ\'e \`a l'UJF, CNRS, BP 166,
38042 Grenoble Cedex 9, France\\
$^2$Dept. of Chemistry, Univ. of Florida, Gainesville, Florida 32611-7200, USA
}

\date{\today}

\begin{abstract}
A method, based on the N\'eel--Brown model 
of thermally activated magnetization reversal of a magnetic single-domain 
particle, is proposed to study the field sweep rate dependence 
of the coercive field of single-molecule magnets (SMMs).
The application to Mn$_{12}$ and Mn$_{84}$ SMMs allows the
determination of the important parameters that characterize
the magnetic properties: the energy barrier, the 
magnetic anisotropy constant, the spin, $\tau_0$, and the crossover
temperature from the classical to the quantum regime.
The method may be particularly valuable for large SMMs that
do not show quantum tunneling steps in the hysteresis loops.
\end{abstract}


%\keywords{}

\maketitle

%%%%%%%%%%%%%%%%%%%%%%%%%%%%%%%%%%%%%%%%%%
\section{Introduction}
\label{intro} 
%%%%%%%%%%%%%%%%%%%%%%%%%%%%%%%%%%%%%%%%%%

Single-molecule magnets (SMMs) 
exhibit slow magnetization relaxation of their spin ground state, 
which is split by axial zero-field 
splitting~\cite{Sessoli93b,Sessoli93,Christou05,Aliaga04,Chakov05}. 
They are now
among the most promising candidates for observing 
the limits between classical and 
quantum physics since they have a well defined 
structure, spin ground state 
and magnetic anisotropy.
An important effort in synthetic chemistry
has led to a quickly growing number of SMMs
with an increasing number of magnetic centers.
Recently, the molecular (or bottom-up)
approach has reached the size regime of the classical (or top-down)
approach to nanoscale magnetic materials~\cite{Tasiopoulos04}.
Indeed, a giant Mn$_{84}$ SMM was reported with a 4 nm diameter
torus structure, exhibiting both magnetization hysteresis and
quantum tunneling.

The study of such large systems is greatly complicated by the fact
that the spin Hilbert space is huge and it is impossible
to treat such systems with exact matrix diagonalization methods.
However, since some SMMs are now as large as some classical
nanoparticles, it raises the interesting possibility
that classical models commonly employed to study the 
latter may be used to obtain
a first-order understanding for large molecular systems. 
Indeed, we herein propose and demonstrate the use of the classical 
N\'eel--Brown model~\cite{Neel49a,Brown63b,Coffey95}
of thermally activated magnetization reversal of a magnetic single-domain 
particle in order to study large SMMs. The proposed method allows
us to determine important parameters that characterize
the magnetic properties of the SMM: the energy barrier, the 
magnetic anisotropy constant, the spin, $\tau_0$, and the crossover
temperature from the classical to the quantum regime.
The method is particularly useful for SMMs
having low-lying energy states and 
not showing quantum tunneling steps in hysteresis loops.
In such systems electron paramagnetic resonance 
(EPR) measurements often exhibit only very broad absorption peaks
which do not allow the determination of
the magnetic anisotropy.

In this letter, we apply the method to two systems
whose properties are already known: (i) a Mn$_{12}$ SMM
with a well-characterized $S=10$ ground state and (ii) a giant Mn$_{84}$ SMM
with $S=6$ which has many low-lying energy states with higher spin values.

%%%%%%%%%%%%%%%%%%%%%%%%%%%%%%%%%%%%%%%%%%
\section{N\'eel--Brown model}
\label{Neel_Brown} 
%%%%%%%%%%%%%%%%%%%%%%%%%%%%%%%%%%%%%%%%%%

The presented method is based on the N\'eel--Brown model 
of thermally activated magnetization reversal of a magnetic single-domain 
particle which has two equivalent ground states of opposite 
magnetization separated by an energy barrier due to 
magnetic anisotropy~\cite{Neel49a,Brown63b,Coffey95}. 
The system can escape from one state to the other either by 
thermal activation over the barrier at high temperatures 
or by quantum tunneling at low temperatures. 
At sufficiently low temperatures and at zero field, 
the energy barrier between the two states of opposite 
magnetization is much too high to observe an escape process. 
However, the barrier can be lowered by applying a magnetic field 
in the opposite direction to that of the particle's magnetization. 
When the applied field is close enough to the reversal field, thermal fluctuations 
are sufficient to allow the system to overcome the barrier, 
and the magnetization is reversed. 

This stochastic escape process can be studied via the 
relaxation time method consisting of the measurement of the
probability that the magnetization has not 
reversed after a certain time. In the case
of an assembly of identical and isolated
particles, it corresponds to measurements
of the relaxation of magnetization.
According to the  N\'eel--Brown model, the probability 
that the magnetization has not 
reversed after a time $t$ is given by:
\begin{equation}
	P(t) = e^{-t/\tau}
\label{eq_P_t}
\end{equation}
and $\tau$ (inverse of the reversal rate) can be expressed by an 
Arrhenius law of the form:
\begin{equation}
	\tau(T,H) = \tau_0 e^{\Delta E(H)/k_{\rm B}T}
\label{eq_tau}
\end{equation}
where $\Delta E(H)$ is the field dependent energy barrier height
and $\tau_0$ is the inverse of the attempt frequency.
In most cases, $\Delta E(H)$ can be approximated by
\begin{equation}
	\Delta E(H) = E_0 \left(1 - H/H_{\rm c}^0\right)^\alpha
\label{eq_E}
\end{equation}
where $H_{\rm c}^0$ is the reversal field at zero temperature,
$E_0$ is the barrier height at zero applied field, and
$\alpha$ is a constant of the order of unity 
(for most cases $1.5 \leq \alpha \leq 2$).
In the case of a Stoner-Wohlfarth particle~\cite{Neel47,St_W48} with
uniaxial anisotropy and the field applied along the easy
axis of magnetization, all constants can be determined
analytically~\cite{Neel47,Neel49a}: $\alpha = 2$, $E_0 = KV$,
and $H_{\rm c}^0 = 2 K/M_{\rm s}$, where $K$ is the 
uniaxial anisotropy constant, $V$ is the particle
volume, and $M_{\rm s}$ is the saturation magnetization.
For SMMs with dominating uniaxial anisotropy:
$\alpha = 2$, $E_0 = DS^2$,
and $H_{\rm c}^0 = 2DS/g\mu_0\mu_{\rm B}$.
However, in general, all constants depend slightly
on fine details of the magnetic anisotropy and
the direction of the applied field $H$~\cite{Thiaville98,Thiaville00}. 

\begin{figure}
\begin{center}
\includegraphics[width=.45\textwidth]{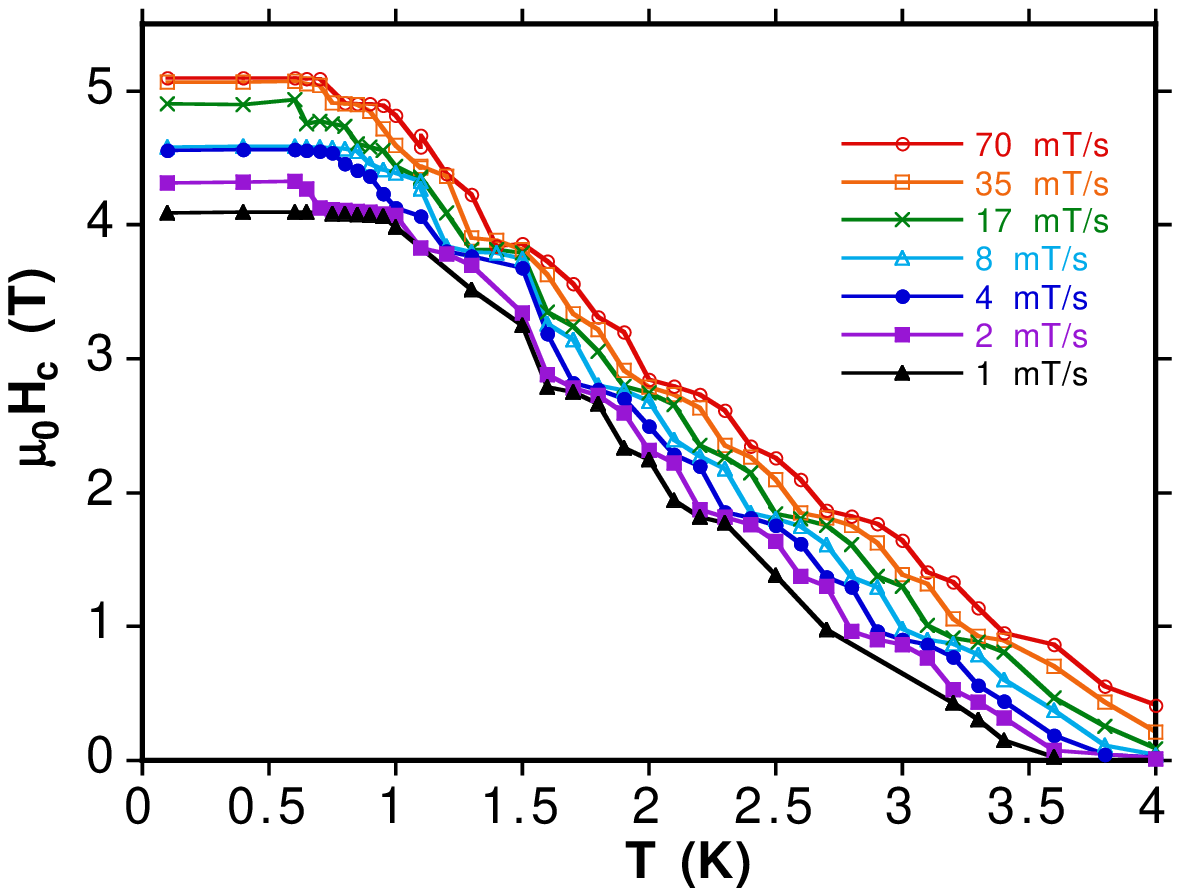}
\includegraphics[width=.45\textwidth]{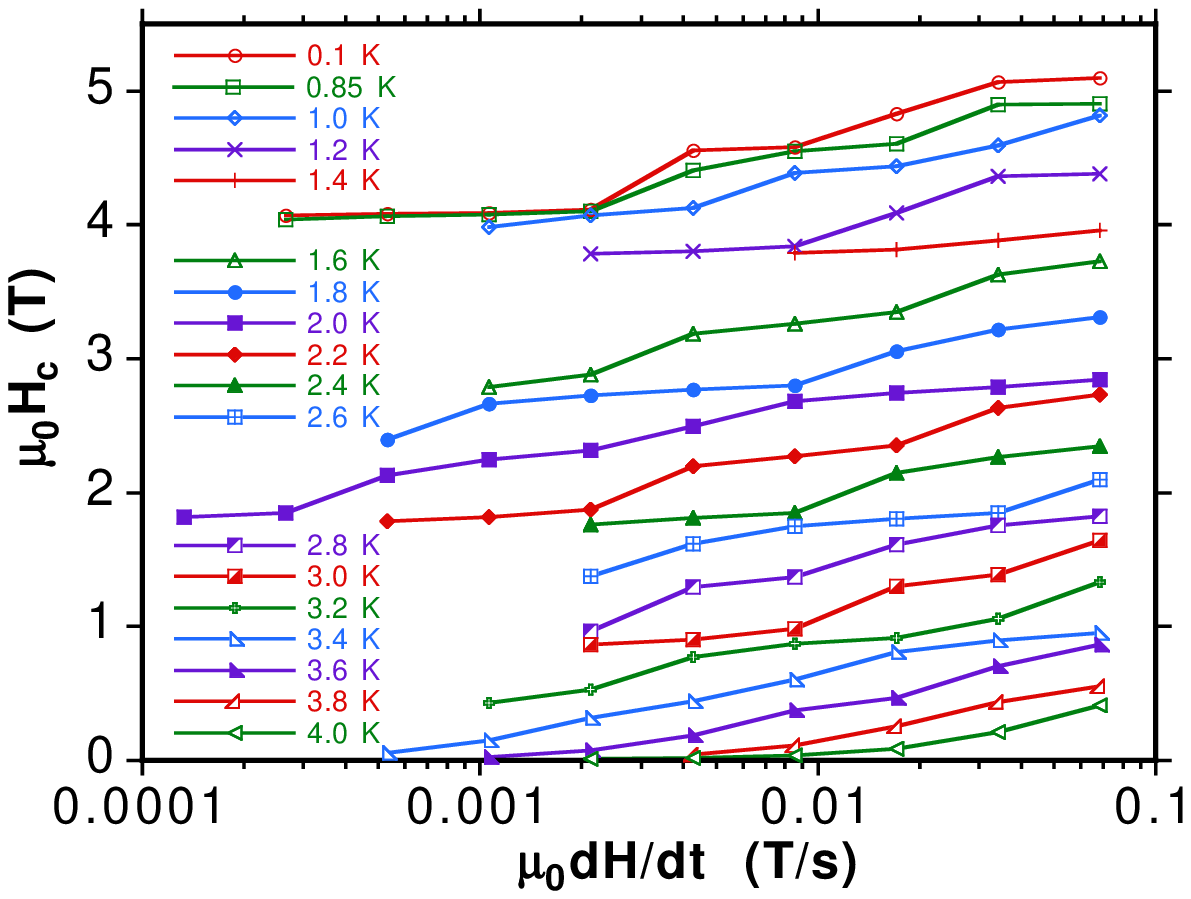}
\caption{(color online) Coercive field $H_{\rm c}$ for Mn$_{12}$
as a function of (a) temperature and (b) field sweep rate.
Note the steps of $H_{\rm c}$ coming from the resonant tunneling 
steps in the hysteresis loops.}
\label{Hc_Mn12}
\end{center}
\end{figure}

\begin{figure}
\begin{center}
\includegraphics[width=.45\textwidth]{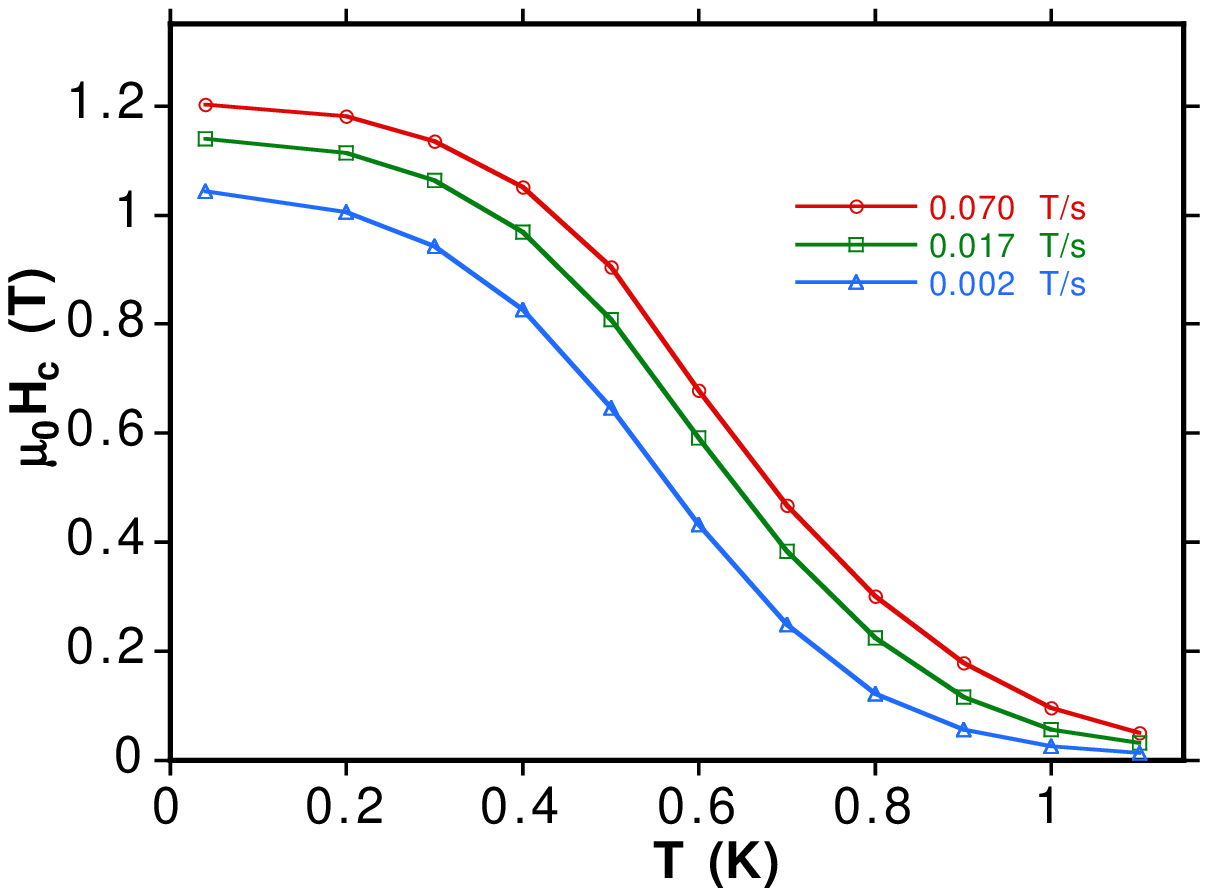}
\includegraphics[width=.45\textwidth]{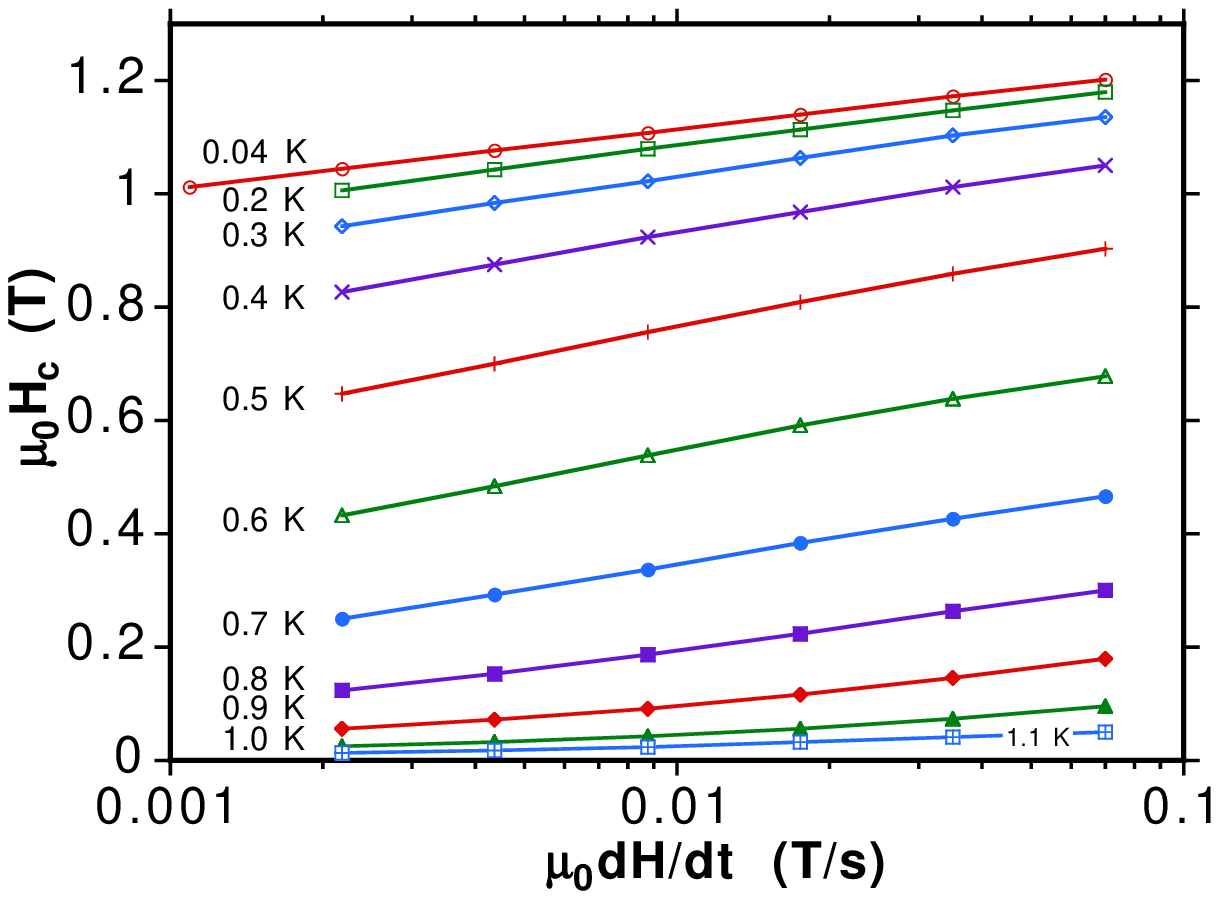}
\caption{(color online) Coercive field $H_{\rm c}$ for Mn$_{84}$
as a function of (a) temperature and (b) field sweep rate.}
\label{Hc_Mn84}
\end{center}
\end{figure}

In order to study the field dependence of the relaxation
time $\tau(T,H)$ and to obtain the parameters
of the model,
the decay of magnetization has to be studied at many
applied fields $H$ and temperatures $T$. This is experimentally
very time consuming and complicated by the fact that
the equilibrium magnetization is temperature dependent
and difficult to obtain for long relaxation times.
In addition, for fast relaxation times the initial magnetization
depends on the field sweep rates to apply the field.
The number of exploitable decades for $\tau$ values is 
therefore limited for relaxation time measurements.

A more convenient method for studying the magnetization decay 
is by ramping the applied field at a 
given rate~\cite{WW_PRL97_Co} and measuring 
the coercive field $H_{\rm c}$ (the field value 
to obtain zero magnetization),
which is then measured as a function of 
the field sweep rate and temperature.

The mathematical transformation from a reversal time probability 
(Eqs.~\ref{eq_P_t} and \ref{eq_tau})
to a reversal field probability was first given by 
Kurkij$\ddot{\rm a}$rvi~\cite{Kurkijarvi72} 
for the critical current in SQUIDs.
Later, Gunther and Barbara calculated
similar expressions for magnetic domain wall 
junctions~\cite{Gunther94}.
A more general calculation was evaluated by Garg~\cite{Garg95}. 
Here, we use a simplified version (see annex)~\cite{WW_PRL97_Co} and
approximate the mean reversal field of an
assembly of identical particles or SMMs by the coercive field $H_{\rm c}$:
\begin{equation}
H_{\rm c}(T,v) \approx H_{\rm c}^0 \left(
	1 - \left\lbrack
	\frac {kT}{E_0} 
	\ln\left(\frac {c}{v}\right)
	\right\rbrack^{1/\alpha}
\right) 
\label{eq_Hc}
\end{equation}
where the field sweeping rate is given by $v = dH/dt$; 
$H_{\rm c}^0$ is the coercive field at zero temperature, and
$c$ depends on the details of the approximations:
$c = H_{\rm c}^0k_{\rm B}T/[\tau_0\alpha E_0(1-H_{\rm c}/H_{\rm 
c}^0)^{\alpha-1}]$
in reference~\cite{WW_PRL97_Co}, 
$c' = H_{\rm c}^0(E_0/kT)^{1/\alpha}/(\tau_0\alpha)$
in reference~\cite{Garg95}, and it can be taken constant
when the exact value of $\tau_0$ is not needed.
We applied the three approximations to nanoparticles~\cite{WW_PRL97_Co}
and here to SMMs and found that the first approximation gives a
$\tau_0$ which is closest to that extracted from an
Arrhenius plot.

%%%%%%%%%%%%%%%%%%%%%%%%%%%%%%%%%%%%%%%%%%
\section{Application to Mn$_{12}$ and Mn$_{84}$}
\label{structure} 
%%%%%%%%%%%%%%%%%%%%%%%%%%%%%%%%%%%%%%%%%%

In the present work, the method is applied to two SMMs: 
(i) ${\rm [Mn_{12}O_{12}(O_2CCH_2Bu^{\it t})_{16}(CH_3OH)_4]\cdot CH_3OH}$ 
and

(ii) ${\rm [Mn_{84}O_{72}(O_{2}CMe)_{78}(OMe)_{24}(MeOH)_{12}
(H_{2}O)_{42}(OH)_{6}]}$,
called respectively Mn$_{12}$ and Mn$_{84}$ henceforth. 
Full details of the synthesis, 
crystal structure and magnetic characterization 
are presented elsewhere~\cite{Tasiopoulos04,Murugesu05}.

The magnetization measurements were 
performed by using (i) a magnetometer consisting
of several 6~$\times$~6~$\mu$m$^{2}$ Hall-bars~\cite{Sorace03}
and (ii) an array of micro-SQUIDs
on top of which single crystals of 
Mn$_{12}$ and Mn$_{84}$ were placed, respectively.
The field can be applied in any direction by separately 
driving three orthogonal superconducting coils. 
The field was aligned with the easy axis of magnetization using
the transverse field method~\cite{WW_PRB04}.

Typical hysteresis loops of both systems can be found elsewhere.
Mn$_{12}$ displays hysteresis loops with a series of quantum steps 
separated by plateaus~\cite{WW_Mn12tBuAc} whereas Mn$_{84}$ shows a smooth 
hysteresis without steps~\cite{Tasiopoulos04}. In order to apply the above
method, the temperature and field sweep rate dependences
of the coercive fields $H_{\rm c}$ were measured and
plotted in Figs~\ref{Hc_Mn12} and~\ref{Hc_Mn84}.
Note that the curves for Mn$_{12}$ show steps coming from the
steps in the hysteresis loops~\cite{WW_Mn12tBuAc}.
As expected for a thermally
activated process, $H_{\rm c}$ increases
with decreasing temperature and increasing field sweep
rate. Furthermore, all our measurements showed an
almost logarithmic dependence of $H_{\rm c}$ on the field sweep
rate (Figs.~\ref{Hc_Mn12}b and~\ref{Hc_Mn84}b).
$H_{\rm c}$ becomes temperature independent below about
0.6 and 0.3 K, respectively for Mn$_{12}$ and Mn$_{84}$.

\begin{figure}
\begin{center}
\includegraphics[width=.45\textwidth]{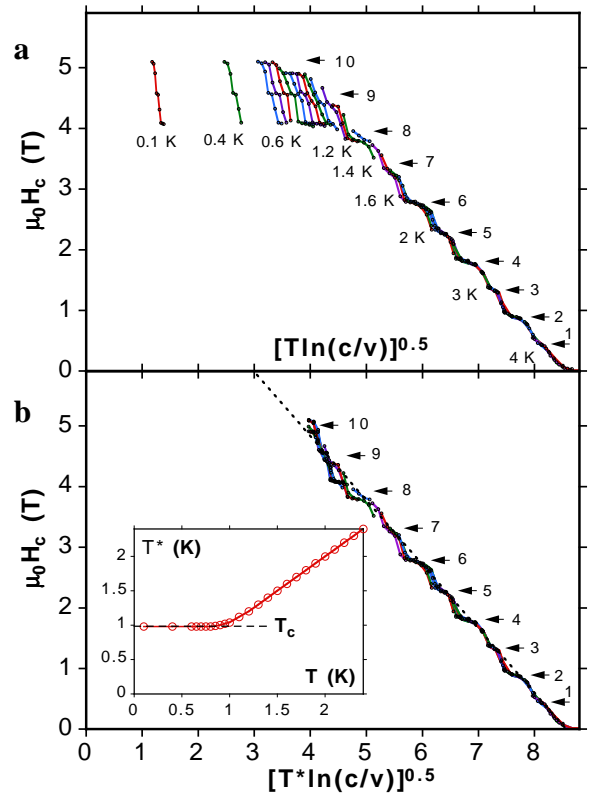}
\caption{(color online) (a) Scaling plot of the coercive field
$H_{\rm c}(T,v)$ of Mn$_{12}$ for field sweep rates
between 0.0001 and 0.1 T/s and several temperatures: 
0.1 K, 0.4 K, from 0.6 to 1 K in steps of 0.05 K,
and from 1 to 4 K in steps of 0.1 K. The arrows indicate
the step index $n = -(m+m')$ where $m$ and $m'$ are the
quantum numbers of the corresponding level crossing.
Note the parity effect of the steps: even $n$ have larger
steps than odd $n$.
(b) Same data of $H_{\rm c}(T,v)$ and same scales but the
real temperature $T$ is replaced by an effective temperature $T^*$ 
(see inset)
which restores the scaling below 1.1 K.}
\label{scaling_Mn12}
\end{center}
\end{figure}

\begin{figure}
\begin{center}
\includegraphics[width=.45\textwidth]{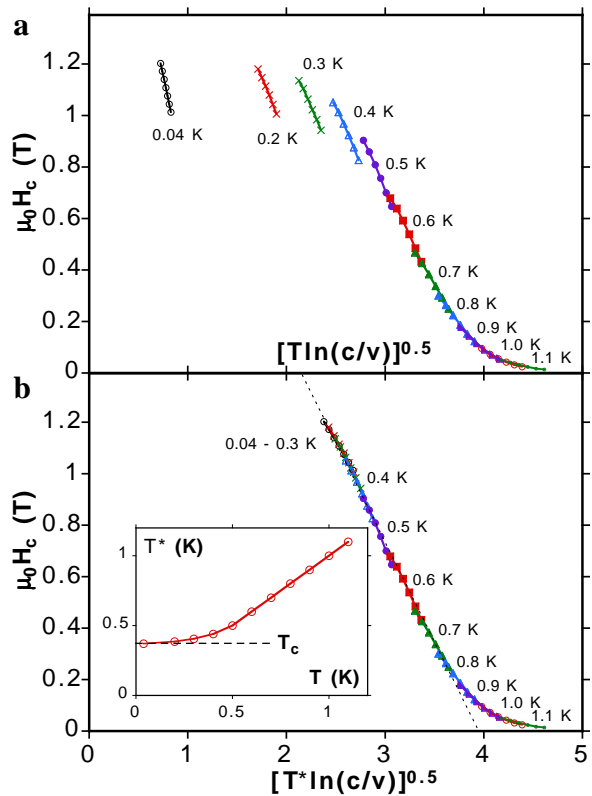}
\caption{(color online) (a) Scaling plot of the coercive field
$H_{\rm c}(T,v)$ of Mn$_{84}$ for field sweep rates
between 0.001 and 0.1 T/s and several temperatures (Fig.~\ref{Hc_Mn84}. 
(b) Same data of $H_{\rm c}(T,v)$ but the
real temperature $T$ is replaced by an effective temperature $T^*$ 
(see inset) which restores the scaling below 0.5 K.}
\label{scaling_Mn84}
\end{center}
\end{figure}

The validity of Eq.~\ref{eq_Hc} was tested by
plotting the set of $H_{\rm c}(T,v)$ values as a function of
$[Tln(c/v)]^{1/2}$ where 
$c = H_{\rm c}^0k_{\rm B}T/\tau_0 2 E_0(1-H_{\rm c}/H_{\rm c}^0)$.
If the underlying model is sufficient, 
all points should collapse onto one straight line by
choosing the proper values for the constant $\tau_0$. 
We found that the data of $H_{\rm c}(T,v)$ fell on a master curve
provided $\tau_0 = 2.1\times 10^{-7}$~s in Fig.~\ref{scaling_Mn12}
and $2\times 10^{-7}$~s in Fig.~\ref{scaling_Mn84}.
Whereas for Mn$_{84}$ the master curve is straight,
it presents steps for Mn$_{12}$.

At low temperatures, strong deviation from the master
curves are observed. 
In order to investigate the possibility that these 
low-temperature deviations are due to escape 
from the metastable potential well by tunneling, a common method 
for classical models is 
to replace the real temperature $T$ by an effective temperature 
$T^*(T)$ in order to restore the scaling plot~\cite{WW_PRL97_BaFeO}. 
In the case of tunneling, $T^*(T)$ should saturate at low temperatures. 
Indeed, the ansatz of $T^*(T)$ as shown in the inset 
of Figs.~\ref{scaling_Mn12}b and~\ref{scaling_Mn84}b, can restore unequivocally 
the scaling plot demonstrated by a straight master 
curve (Figs.~\ref{scaling_Mn12}b and~\ref{scaling_Mn84}b). 
The flattening of $T^*$ corresponds to a saturation 
of the escape rate, which is a necessary signature of 
tunneling. 
The crossover temperature $T_{\rm c}$ can be defined as 
the temperature where the quantum 
rate equals the thermal one. 
The inset 
of Figs.~\ref{scaling_Mn12}b and~\ref{scaling_Mn84}b
gives $T_{\rm c}$ = 0.97 and 0.37 K for Mn$_{12}$ and Mn$_{84}$, respectively.
The slopes and the intercepts of the master curves give 
$E_0$ = 72.4 and 15.6 K and $H_{\rm c}^0$ = 9.1 and 3.1 T,
respectively for Mn$_{12}$ and Mn$_{84}$.
The $E_0$ values are in good agreement with those extracted
from Arrhenius plots (69 K and 18 K for 
Mn$_{12}$ and Mn$_{84}$, respectively)~\cite{Murugesu05,Tasiopoulos04}.
This result allows us to estimate the spin ground
state using $S = 2E_0/(g\mu_{\rm B}\mu_0H_{\rm c}^0)$:
$S$ = 11 and 7 for Mn$_{12}$ and Mn$_{84}$, respectively.
This differs slightly from $S$ = 10 and 6 determined via
magnetization measurements. This deviation is due to quantum effects
in the thermally activated regime and is considered further below.

Several points should be mentioned:
(i) the classical regime of the model corresponds in most SMMs to
the thermally activated tunneling regime with
tunneling close to the top of the energy barrier.
Because all parameters are deduced from this regime,
small deviations from the exact values are expected;
(ii) Eq.~\ref{eq_Hc} is not valid for fields which
are close to $H=0$ because the model only takes into account
the transitions from the metastable to the stable well.
However, close to $H=0$, transitions between both wells are possible
leading to a rounding of the master curve at small fields;
(iii) the method can be applied to powder samples with random
orientations of the molecules. In this case, $\alpha \approx 1.5$,
$\nu E_0 = DS^2$ where $\nu$ can be 
calculated~\cite{Thiaville98,Thiaville00},
and the intercept of the master curve gives $H_{\rm c}^0/2$;
(iv) in the case of a distribution of anisotropies, different
parts of the distribution can be probed by applying the 
method at different $M$ values;
(v) this method is insensitive to small intermolecular
interactions when $H_{\rm c}$ is larger than the typical
interaction field; and
(vi) the method can be generalized for 1D, 2D, and 3D networks
of spins. In this case, Eq.~\ref{eq_E} describes a nucleation barrier.

\begin{figure}
\begin{center}
\includegraphics[width=.45\textwidth]{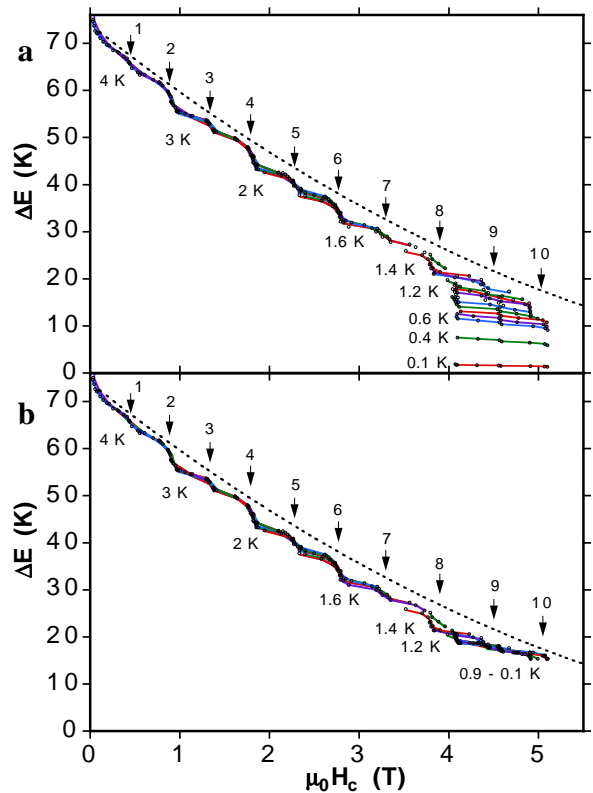}
\caption{(color online) (a) Field dependence of the
energy barrier of Mn$_{12}$ obtained from Eq.~\ref{eq_barrier}
and the set of $H_{\rm c}(T,v)$ data from Fig.~\ref{scaling_Mn12}. 
The arrows indicate
the step index $n = -(m+m')$ where $m$ and $m'$ are the
quantum numbers of the corresponding level crossing.
Note the step like reduction of the 
energy barrier due to resonant tunneling
and the parity effect of the steps: even $n$ have larger
steps than odd $n$.
The dotted line gives the classical barrier
$\Delta E = E_0(1-H/H_{\rm a})^2$ with $E_0$ = 74 K and 
$H_{\rm a}$ = 9.8 T.
(b) Same data of $H_{\rm c}(T,v)$ but the
real temperature $T$ is replaced by an effective temperature $T^*$ 
(see inset of Fig.~\ref{scaling_Mn12}b).}
\label{barrier_Mn12}
\end{center}
\end{figure}

\begin{figure}
\begin{center}
\includegraphics[width=.45\textwidth]{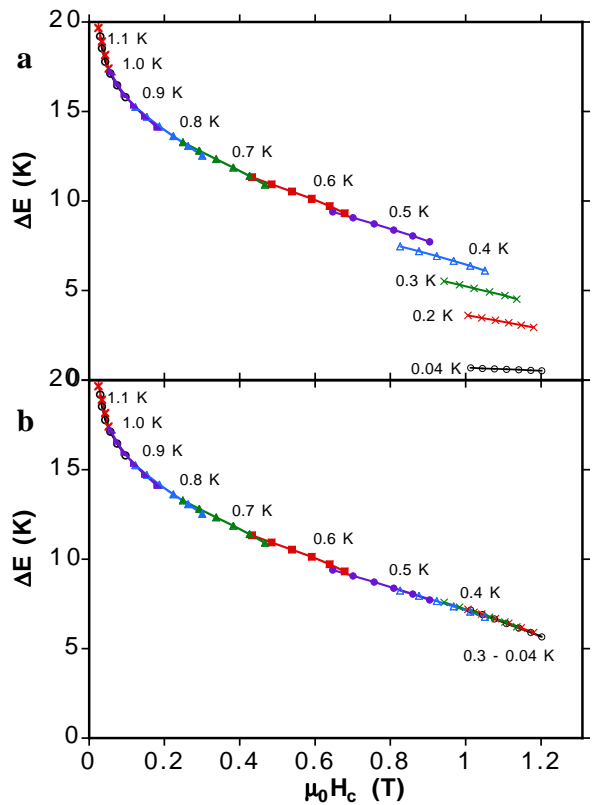}
\caption{(color online) (a) Field dependence of the
energy barrier of Mn$_{84}$ obtained from Eq.~\ref{eq_barrier}
and the set of $H_{\rm c}(T,v)$ data from Fig.~\ref{scaling_Mn84}. 
(b) Same data of $H_{\rm c}(T,v)$ but the
real temperature $T$ is replaced by an effective temperature $T^*$ 
(see inset of Fig.~\ref{scaling_Mn84}b).}
\label{barrier_Mn84}
\end{center}
\end{figure}

%%%%%%%%%%%%%%%%%%%%%%%%%%%%%%%%%%%%%%%%%%
\section{Conclusion}
%%%%%%%%%%%%%%%%%%%%%%%%%%%%%%%%%%%%%%%%%%
We introduce a method that uses the
temperature and field sweep rate dependences of
the coercive field of large SMMs in order to determine several
parameters that characterize
the magnetic anisotropy and the relaxation dynamics.
We have succesfully applied the method to two test cases:
(i) a Mn$_{12}$ SMM
with well known magnetic properties; and (ii) a giant Mn$_{84}$ SMM
which has many low-lying energy states with higher spin values.
We believe that this method is an important tool to characterize
magnetically new molecular systems of great complexity
which do not allow a detailed undestanding on the quantum level.

This work was supported by the EC-TMR Network 
QuEMolNa (MRTN-CT-2003-504880), CNRS, 
Rh${\rm\hat{o}}$ne-Alpes funding,
and NSF. 

%%%%%%%%%%%%%%%%%%%%%%%%%%%%%%%%%%%%%%%%%%
\section{Annex}
%%%%%%%%%%%%%%%%%%%%%%%%%%%%%%%%%%%%%%%%%%

The probability density of reversal of a stochastic process is
\begin{equation}
    -\frac{dP}{dt} = \frac{1}{\tau}P
\label{eq_density}
\end{equation}
and the maximum of the probability density can be
derived from
\begin{equation}
    \frac{d^2P}{dt^2} =  
    \frac{P}{\tau^2}\left(1+\frac{d\tau}{dt}\right) = 0
\label{eq_d2P}
\end{equation}
This gives the following general result for the 
maximum of the probability density
\begin{equation}
    \frac{d\tau}{dt} = -1
\label{eq_max_density}
\end{equation}
The application of the result to Eq.~\ref{eq_tau} leads to
\begin{equation}
    \Delta E(H) = k_{\rm B}T{\rm ln}\left(\frac{k_{\rm 
    B}T}{\tau_0\frac{dE}{dH}\frac{dH}{dt}}\right)
\label{eq_barrier}
\end{equation}
Using Eqs.~\ref{eq_E} and~\ref{eq_barrier}, we find  Eq.~\ref{eq_Hc}.

Eq.~\ref{eq_barrier} can be used to plot directly the
field dependence of the energy barrier (Figs.~\ref{barrier_Mn12} 
and~\ref{barrier_Mn84})

%\bibliographystyle{wernsdor} 
%\bibliography{wernsdor}

\begin{thebibliography}{10}

\bibitem{Sessoli93b}
R. Sessoli, H.-L. Tsai, A.~R. Schake, S. Wang, J.~B. Vincent, K. Folting, D.
  Gatteschi, G. Christou, and D.~N. Hendrickson, J. Am. Chem. Soc. {\bf 115},
  1804  (1993).

\bibitem{Sessoli93}
R. Sessoli, D. Gatteschi, A. Caneschi, and M.~A. Novak, Nature {\bf 365},  141
  (1993).

\bibitem{Christou05}
G. Christou, Polyhedron XX, XXX  (2005).

\bibitem{Aliaga04}
N. Aliaga-Alcade, R.S. Edwards, S.O. Hill, W. Wernsdorfer, K. Folting and G. Christou, 
J. Am. Chem. Soc. 126, 12503  (2004).

\bibitem{Chakov05}
N. E. Chakov, M. Soler, W. Wernsdorfer, K.A. Abboud, and G. Christou, Inorg. Chem. 44, 5304  (2005).

\bibitem{Tasiopoulos04}
A.J. Tasiopoulos, A. Vinslava, W. Wernsdorfer, K.A. Abboud, and G. Christou,
  Angew. Chem. Int. Ed. Engl. {\bf 43},  2117  (2004).

\bibitem{Neel49a}
L. N\'eel, Ann. Geophys. {\bf 5},  99  (1949).

\bibitem{Brown63b}
W.~F. Brown, Phys. Rev. {\bf 130},  1677  (1963).

\bibitem{Coffey95}
W.~T. Coffey, D.~S.~F. Crothers, J.~L. Dormann, Yu.~P. Kalmykov, , and J.~T.
  Waldron, Phys. Rev. B {\bf 52},  15951  (1995).

\bibitem{Neel47}
L. N\'eel, C. R. Acad. Science {\bf 224},  1550  (1947).

\bibitem{St_W48}
E.~C. Stoner and E.~P. Wohlfarth, Philos. Trans. London Ser. A {\bf 240},  599
  (1948), reprinted in IEEE Trans. Magn. MAG-27, 3475 (1991).

\bibitem{Thiaville98}
A. Thiaville, J. Magn. Magn. Mat. {\bf 182},  5  (1998).

\bibitem{Thiaville00}
A. Thiaville, Phys. Rev. B {\bf 61},  12221  (2000).

\bibitem{WW_PRL97_Co}
W. Wernsdorfer, E.~Bonet Orozco, K. Hasselbach, A.~Benoit~B. Barbara, N.
  Demoncy, A. Loiseau, D. Boivin, H. Pascard, and D. Mailly, Phys. Rev. Lett.
  {\bf 78},  1791  (1997).

\bibitem{Kurkijarvi72}
J. Kurkij$\ddot{\rm a}$rvi, Phys. Rev. B {\bf 6},  832  (1972).

\bibitem{Gunther94}
L. Gunther and B. Barbara, Phys. Rev. B {\bf 49},  3926  (1994).

\bibitem{Garg95}
A. Garg, Phys. Rev. B {\bf 51},  15592  (1995).

\bibitem{Murugesu05}
M. Murugesu and et~al., to be submitted  (2005).

\bibitem{Sorace03}
L. Sorace, W. Wernsdorfer, C. Thirion, A.-L. Barra, M. Pacchioni, D. Mailly,
  and B. Barbara, Phys. Rev. B {\bf 68},  220407(R)  (2003).

\bibitem{WW_PRB04}
W. Wernsdorfer, N.~E. Chakov, and G. Christou, Phys. Rev. B {\bf 70},  132413
  (2004).

\bibitem{WW_Mn12tBuAc}
W. Wernsdorfer, M. Murugesu, and G. Christou, cond-mat/0508437  .

\bibitem{WW_PRL97_BaFeO}
W. Wernsdorfer, E.~Bonet Orozco, K. Hasselbach, A. Benoit, D. Mailly, O. Kubo,
  H. Nakano, and B. Barbara, Phys. Rev. Lett. {\bf 79},  4014  (1997).

\end{thebibliography}

\end{document}